# Spin-orbit coupling in single layer ferrimagnets: direct observation of spin-orbit torques and chiral spin textures

Sachin Krishnia[1,*,†], Eloi Haltz[1,*], Léo Berges[1], Lucia Aballe[2], Michael Foerster[2], Laura Bocher[1], Raphaël Weil[1], André Thiaville[1], João Sampaio[1‡], and Alexandra Mougin[1]

[1]Université Paris-Saclay, CNRS, Laboratoire de Physique des Solides, 91405, Orsay, France.

[2]Alba Synchrotron Light Facility, CELLS, Barcelona, E-08290, Spain.

**Abstract**

We demonstrate that effects of spin-orbit coupling and inversion asymmetry exist in a single GdFeCo ferrimagnetic layer, even without a heavy metal interface. We use electric transport measurements to quantify the spin-orbit torques. We measure the Dzyaloshinskii-Moriya interaction using Brillouin light scattering measurement technique, and we observe the resulting chiral magnetic textures using x-ray PEEM microscopy. We attribute these effects to a composition variation along the thickness, that we observed by scanning transmission electron microscopy. We show that these effects can be optimized by varying the GdFeCo thickness or by combining them with interfacial effects.

## I. Introduction

Magnetic layers interfaced with materials with strong spin-orbit coupling (SOC) (such as Pt, Ta) manifest several fascinating phenomena when the inversion symmetry is broken [1–4]. One direct consequence of SOC is the interplay between charge and spin transport *via* spin Hall effect (SHE) in heavy metals (HM), which leads to non-equilibrium spin accumulation at the surfaces [5,6]. Another mechanism, the Rashba effect, arises when the electrical carriers move in an interfacial electric field and experience the resultant magnetic field that couples with their spins [7,8]. Both mechanisms give rise to spin-orbit torques (SOTs) in the adjacent magnetic layer, with damping-like and field-like components [9–11]. Surprisingly, SOTs have been detected in a single layer NiFe even in the absence of HM adjacent layers [12,13]. Moreover, SOTs are an efficient way to manipulate chiral magnetic textures, such as Néel domain walls (DWs) [2,3] and skyrmions [14], whose chirality is induced by another product of SOC, the Dzyaloshinskii-Moriya interaction (DMI) [1].

---

† sachinbudana@gmail.com

‡ joao.sampaio@universite-paris-saclay.fr

* These authors have equal contribution

Recently, the combination of SOTs and DMI has enabled fast and efficient magnetization dynamics in rare earth (RE)-transition metal (TM) ferrimagnetic films interfaced with a HM [15–17]. In conventional 3d-TM/HM bilayers, Pt 5d (or Pd 4d) states hybridize strongly with the 3d states of the TM leading to proximity effects and an enhanced orbital moment [8,18]. In RE-TM alloys, spintronic effects are described by the combination of localized 4f magnetism and itinerant magnetism of the *spd* band structure [19]. It was shown earlier that the itinerant magnetism is dominated by RE 5d and TM 3d states [18]. This hybridization implies that RE atoms with unpaired 5d electrons may exhibit large SOC [20–22], suggesting that SOT and DMI could appear even in the absence of an adjacent HM layer. Very recent observation of interfacial DMI in Tm and Tb-based oxide garnets is attributed to the orbital moment of the RE [23,24]. Furthermore, an interface-independent DMI originated in the bulk of a GdFeCo ferrimagnet layer has been reported using indirect methods based on asymmetric magnetic reversal [25], which was attributed to a broken symmetry caused by inhomogeneity of RE atoms along the film thickness. On the other hand, experiments on [Co/Tb]$_n$ multilayers did not show any SOT contribution from the RE [26]. The underlying origin of SOC and the role of RE on SOC in ferrimagnets hence remain to be understood and need to be addressed by direct experimental evidence.

In this letter, we provide direct evidence of the internal SOC effects in single layer GdFeCo ferrimagnets. The SOC-induced effects such as SOTs, DMI and chiral magnetic textures, are observed in a GdFeCo film without any HM layer, but with a symmetry-breaking inhomogeneous Gd concentration along the thickness, a key ingredient to observe the SOC effects.

## II. Film Deposition and Characterization

$Gd_{0.35}(Fe_{85}Co_{15})_{0.65}$ 5 nm thick films were deposited on thermally-oxidized Si substrates by e-beam co-evaporation of Gd and FeCo targets in ultra-high vacuum (~$10^{-10}$ mbar) [27]. Al (5 nm) and Pt (7 nm) capping layers were deposited to prevent oxidation of the ferrimagnets. The Pt-capped sample is used as a reference and, hereafter, GdFeCo refers to the Al-capped sample unless mentioned. The GdFeCo film is composed of two sub-lattices of Gd (RE) and FeCo (TM) coupled antiferromagnetically [28]. Magnetization as a function of temperature is shown in Figure 1(a). At the magnetic compensation temperature, $T_M \approx 275K$ for this sample, the net magnetization vanishes and the coercive field diverges (Figure 1b).

To measure the Hall effect, the films were patterned into 5 μm wide Hall cross structures using e-beam lithography and a hard-mask etching technique. Figure 1(c) and (d) show hysteresis loops obtained by anomalous Hall effect below and above $T_M$. The electrical properties are dominated by the TM sub-lattice and, therefore, the change of sign of the Hall resistance ($R_{AHE}$) across $T_M$ is a signature of the reversal of the alignment of FeCo sub-lattice with the external field, as shown in the insets. The effective anisotropy field ($H_k$; Figure 1(b) in blue), obtained by fitting $R_{AHE}$ versus in-plane field [29], diverges at $T_M$. The spin-flop transitions were not observed in the measured temperature and magnetic field range (See Supplemental Material at [30]).

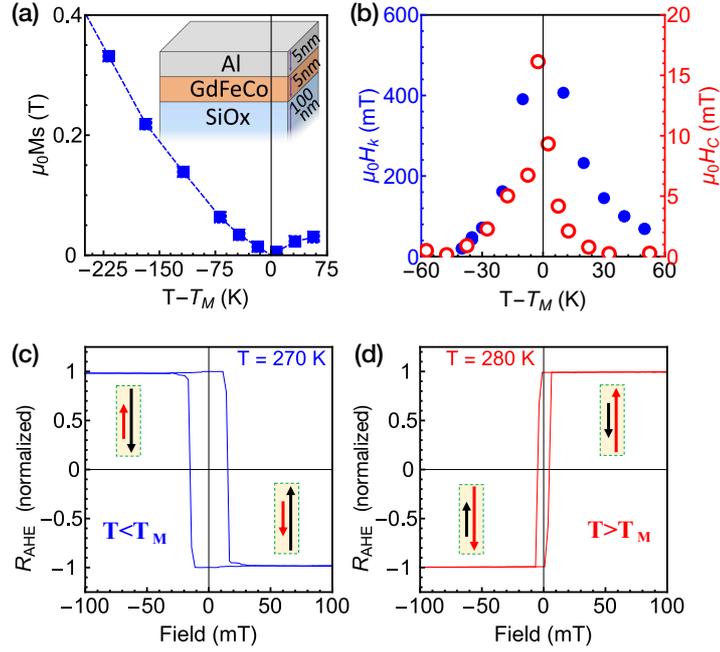

**Figure 1.** (a) Net magnetization ($M_s$) versus temperature measured using SQUID magnetometry. (b) Anisotropy field ($H_k$) in blue and coercive field ($H_c$) in red as a function of temperature, which diverge at $T_M$. (c,d) Normalized $R_{AHE}$ vs perpendicular magnetic field at (c) T = 270 K < $T_M$ and (d) T = 280 K > $T_M$. The arrows represent the direction and relative magnitude of Gd (black) and FeCo (red) magnetic moments.

## III. Current-Induced Torques

To investigate the existence of SOC inside the ferrimagnetic layer, we quantify the two components of current-induced torques by using second harmonic Hall voltage measurement technique [9,31]. The measurement geometries are shown in Figure 2(a,b). This technique uses an ac current of low frequency ($f$ = 1.33 kHz) to induce periodic magnetization oscillations, which modulate the $R_{AHE}$ at frequency $f$ and the anomalous Hall voltage at frequency $2f$. By measuring simultaneously the first and second harmonics as a function of in-plane field (H) applied along or transversely to the current direction (H//$I_{AC}$ and H⊥$I_{AC}$, Figure 2a,b), we extract the damping-like ($H_{DL}$) or field-like ($H_{FL}$) effective fields (See Supplemental Material at [30]). To increase the sensitivity, the analysis method was improved to include magnetic fields larger than $H_k$.

Figure 2(c) shows the temperature dependence of $H_{DL}$ per current density ($H_{DL}/J$). At a given temperature, $H_{DL}$ increases linearly with current (showing that thermally-induced Hall voltage is negligible; See Supplemental Material at [30]) and changes sign when the magnetization is reversed, in agreement with the expected symmetry of SHE induced torque: $\vec{H}_{DL} = H_{DL}(\hat{m} \times \hat{\sigma})$, where $\hat{m}$ and $\hat{\sigma}$ are the unit vectors along magnetization and spin polarization directions [32]. Assuming that SHE is the main source of $H_{DL}$, we expect that $\mu_0 H_{DL} = \frac{\hbar}{2e}\frac{1}{M_s t}J\theta_{SH}$, where $\theta_{SH}$ is the spin Hall angle and should be temperature-independent for a given material. $H_{DL}/J$ diverges at $T \to T_M$, showing the expected scaling with $1/M_S$. We find $\theta_{SH} \approx 7 \times 10^{-4} > 0$ (figure 2e), which is ~100x smaller than the reported Co/Pt interfacial effect ($\theta_{SH} \sim 0.05$ in Co/Pt) [33]. $\theta_{SH}$ is independent of temperature, in particular not changing across $T_M$, which is consistent with the hypothesis of SHE as the main source of $H_{DL}$.

Figure 2(d) shows the $H_{FL}$ per current density ($H_{FL}/J$) at various temperatures below the $T_M$. We could not reliably extract the $H_{FL}$ above the $T_M$ due to increasing noise with temperature (See Supplemental Material at [30]). At a fixed temperature, $H_{FL}$ does not change sign with the reversal of the magnetization, as expected with the symmetry: $\vec{H}_{FL} = H_{FL}\hat{\sigma}$. In addition to SOTs, the current generates an Oersted field ($H_{Oe}$) that is independent of $M_s$ but has the same orientation as $H_{FL}$. Therefore, $H_{Oe}$ contributes to the measured $H_{FL}$ (see a horizontal line in Fig. 2d). However, the divergence of $H_{FL}$ as $T \to T_M$ shows that a non-Oersted torque is present. Assuming a Rashba-induced $H_{FL}$, it would depend on $M_s$ with a relation $\mu_0 H_{FL} = -\frac{2\,m_e}{\hbar e M_s}\alpha_R PJ$, where $m_e$ is the electron mass, $P$ is the current polarization and $\alpha_R$ is the Rashba parameter. This relation matches well with the obtained variation of $H_{FL}$. Figure 2(f) shows the extracted $\alpha_R P$ at various temperatures. The obtained value of $\alpha_R P \approx 7 \times 10^{-33} Jm$ is independent of the temperature, suggesting the presence of significant SOT-induced $H_{FL}$. The existence of these SOTs in a ferrimagnetic layer without an adjacent HM suggests the presence of an internal SOC from the magnetic layer itself.

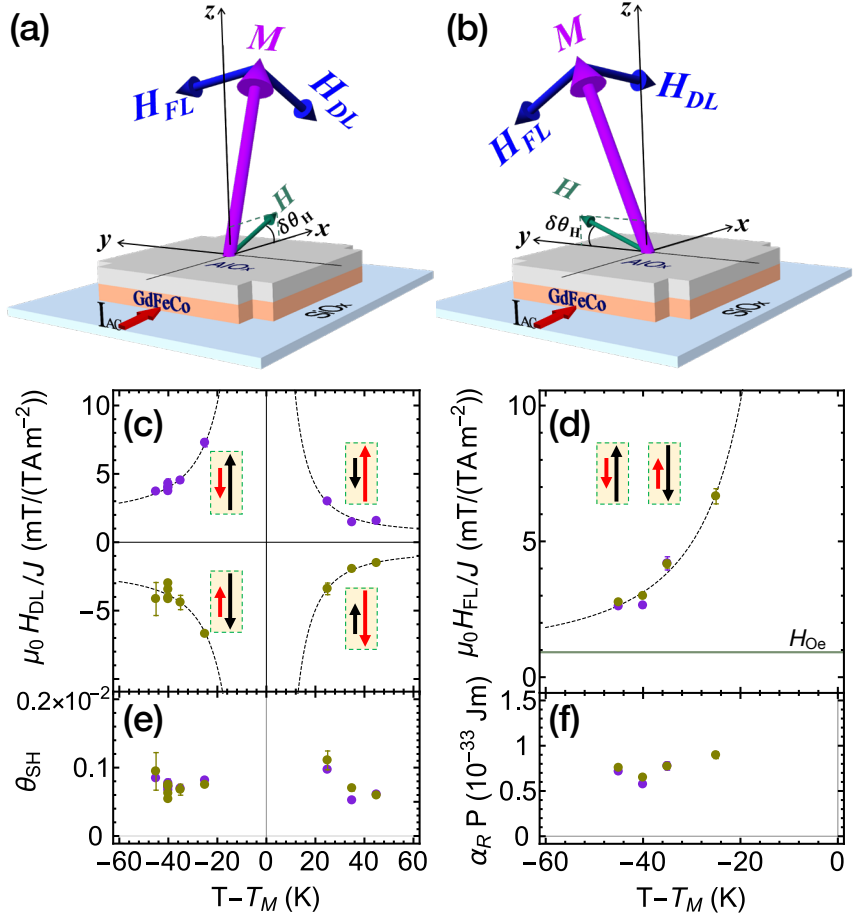

**Figure 2** (**a,b**) Illustration of harmonic Hall measurement in (**a**) longitudinal and (**b**) transverse geometries to obtain the $H_{DL}$ and $H_{FL}$. $H$ and $I_{ac}$ are the applied magnetic field and ac current. (**c**) $H_{DL}/J$ (with $J$ estimated in the GdFeCo layer) and (**d**) $H_{FL}/J$ vs temperature. Purple (khaki) points correspond to an up (down) saturated state. The diagrams represent the direction and magnitude of Gd (black) and FeCo (red) magnetic moments. The dotted lines are guides to the eyes. The estimated Oersted field ($\sim 1 mT/TAm^{-2}$) due to $I_{ac}$ is shown by a line. (**e**) Spin Hall angle ($\theta_{SH}$) and (**f**) Rashba parameter ($\alpha_R P$) vs temperature.

## IV. Chiral Spin Textures and Dzyaloshinskii-Moriya Interaction

Another phenomenon induced by SOC is the DMI, which favors chiral magnetic textures [15,34,35]. We examine the internal structure of DWs in the ferrimagnetic film using photoemission electron microscopy combined with X-ray magnetic circular dichroism (XMCD-PEEM) [36]. In our experiment, the incident X-ray energy was set to the Gd $M_{4,5}$ absorption edge at a grazing angle of 16° (see Fig. 3a). In XMCD, the contrast is proportional to $\hat{k}_{X-ray} \cdot \hat{m}$ (where $\hat{k}_{X-ray}$ is the unit vector along the incident beam), resulting in 3.5x higher in-plane than that of out-of-plane contrast. Therefore, Néel and Bloch DWs, as well as their chirality, can be distinguished by their different XMCD-PEEM profiles [35], even when the DW width ($\Delta$) is smaller than the microscope resolution ($r$) (see Fig. 3b).

The imaging was performed at $T = 230$ K $< T_M$ to avoid thermal magnetization fluctuations. Figure 3(c) shows typical XMCD-PEEM images of up and down domains manifested by light and dark grey contrasts, respectively (See Supplemental Material at [30] for more images). Interestingly, an intense black or white contrast appears between the domains when the DW length is perpendicular to the incident beam, while it is absent when the incident beam is along the DW length. The intense DW contrast is clearly visible on the XMCD line scans (Figure 3d), as evidenced by either a peak or a dip between the dark and light grey domains. This pattern shows that the DW magnetization lies parallel or antiparallel to the X-ray beam, that is, the DW is Néel with left-handed chirality. To acquire the DWs width and internal magnetization direction, we fit the obtained XMCD line scans with the expected profile convoluted with a Gaussian function to account for the microscope resolution. By fitting several line-scans of DWs with different orientations keeping the azimuthal angle ($\varphi$) and $\Delta$ as free parameters (figure 3d), we obtained $\varphi = 180 \pm 10°$ (left-handed Néel) and $\Delta = 20 \pm 10$ nm. As the ferrimagnetic film is not interfaced with any HM, these chiral Néel DWs clearly indicate the existence of DMI in the volume of the film.

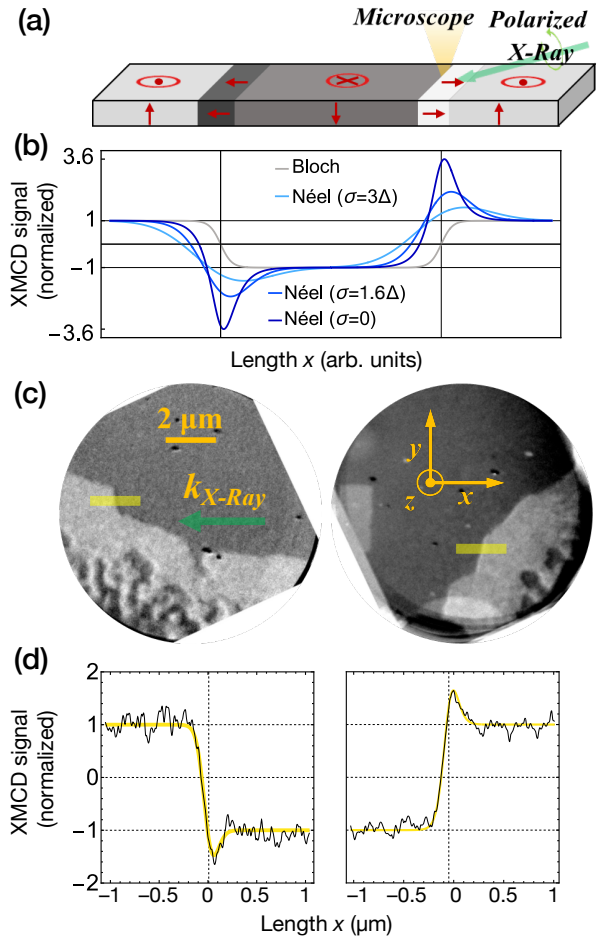

**Figure 3 (a)** Schematic of the XMCD-PEEM experiment, showing the magnetization profile of an up-down and down-up left-handed Néel DWs and the X-ray beam with grazing incidence (16°) and perpendicular to the DW length. **(b)** Calculated normalized XMCD-PEEM intensity profile for DWs in (a), taking into account the finite microscope resolution (*r*; blue curves). The gray curve is the profile of Bloch DWs. **(c)** Multidomain XMCD-PEEM images at the Gd $M_{4,5}$ edge (1178.7 eV). The green arrow shows the direction of the X-ray beam. The dark (bright) contrast corresponds to a down (up) magnetic domain. The analyzed DWs are in the vicinity of a region of lowered anisotropy induced by long exposure to the X-rays (with small domains; in the bottom). **(d)** Intensity profiles (black lines) averaged over the yellow regions in (c). The thick yellow line is a fit using the theoretical profile as shown in (a) with $r = 60$ nm that was measured on dust particles in the image.

To quantify the magnitude and sign of DMI, we perform spin-wave (SW) spectroscopy experiments using Brillouin light scattering (BLS) technique in Damon-Eshbach geometry (see inset in Figure 4a) at room temperature (293 K) [37]. This method measures Stokes ($f_S$) and anti-Stokes ($f_{AS}$) resonance frequencies, which correspond to counter-propagating SWs (Fig. 4a). Due to DMI, they have different frequencies. The variation of this difference $\Delta f = f_{AS} - f_S$ with the incident wave vector $k_{sw}$ is directly proportional to the DMI parameter $D$: $\Delta f = -\frac{2\gamma}{\pi M_s} D k_{sw}$, with $\gamma$ being the gyromagnetic ratio of the alloy [37–42] (See Supplemental Material at [30]).

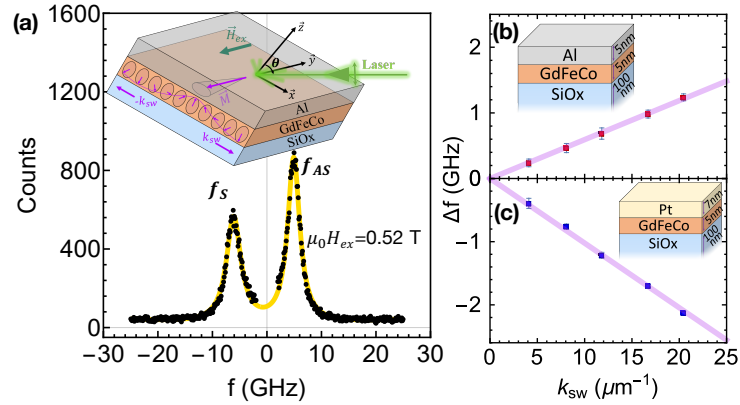

**Figure 4 (a)** BLS spectrum of SiOx/GdFeCo/Al obtained with a laser wavelength of 532 nm at 293 K under an in-plane magnetic field $\mu_0 H_{ex} = 0.52$ T and at beam incidence angle ($\theta$) = 60° (corresponding to $|k_{sw}|$= 20.46 μm$^{-1}$). The yellow lines are Lorentzian fits. The inset shows measurement geometry. **(b,c)** The DMI-induced frequency shifts ($\Delta f$) of nonreciprocal SWs *vs* wave vector magnitudes $|k_{sw}|$ for **(b)** Al-covered (Red) and **(c)** Pt-covered (Blue) film. The thick lines are linear fits. Both samples were measured at T < $T_M$.

A typical BLS spectrum for the same sample is shown in figure 4(a) and the dependency of $\Delta f$ with $k_{sw}$ is shown in Figure 4(b). A value of $D = -8.4 \pm 2.5$ μJ/m$^2$ is obtained from $\Delta f$ vs $k_{sw}$. This DMI magnitude is much smaller than Ta/Co and Pt/Co systems [38].

To understand the stabilization of Néel DWs with such low *D*, we must consider the critical DMI ($D_C = \frac{4}{\pi} \Delta K \approx \frac{2 \ln 2}{\pi^2} t \mu_0 M_S^2$) required to form a Néel DW in thin films [1]. Due to the low $M_S$ of this system, *D* is indeed ~4x larger than the estimated $D_C$ at the temperature of the PEEM experiments ($D_C \approx 2.3$ μJ/m$^2$). This shows that a small DMI can still be dominant in low $M_s$ systems. Therefore, the extracted $D$ favors left-handed ($D < 0$) Néel DW structure ($|D| > D_C$) as observed in the PEEM experiments.

Further insights into the origin of these effects are obtained by comparing the $\theta_{SH}$ and $D$ of this film with a control sample with a Pt capping layer (Si/SiOx/GdFeCo 5 nm /Pt 7 nm), for which the usual interfacial effects should be present. We find a median value of $\theta_{SH} = -0.12$ in the GdFeCo/Pt sample (See Supplemental Material at [30]), with an opposite sign to the GdFeCo/Al. Furthermore, the DMI parameter $D$ for the Pt-capped control sample was determined: $D = (27.2 \pm 0.2)\ \mu J/m^2$ (figure 4c). The sign of $D$ is the same as reported for the Co/Pt interface [38], and is opposite to that of GdFeCo/Al sample. The magnitude of $D$ agrees with the naïve expectation that GdFeCo/Pt, containing only 10% of Co, should show about ~10% of the DMI of pure Co/Pt interfaces [38]. These findings show that DMI and SOT in the control GdFeCo/Pt sample are dominated by the interfacial effects, unlike what was found for DMI in another study [25].

## V. Thickness Dependence of DMI and SOTs

To investigate the origin of DMI and SOTs in GdFeCo/Al, we have studied samples with different GdFeCo thicknesses. The measured DMI values for 4 nm, 5 nm, 6 nm and 8 nm thick samples are shown in figure (5a) and they do not show a variation with thickness. The interfacial mechanism would exhibit 1/t variation of DMI parameter as found in Pt/Co samples [38], where 't' is the thickness of magnetic layer. Therefore, these results suggest that the DMI cannot be produced by a pure interfacial mechanism.

Furthermore, we have measured SOTs in GdFeCo (8nm)/Al(5nm) sample and a five-fold increase in SOTs is observed (figure 5b), which further excludes the possibility of a dominating interfacial SOC mechanism and suggests a bulk origin of DMI and SOTs. However, the interfacial mechanism dominates over bulk contributions when the GdFeCo is interfaced with a strong SOC material as we have shown in GdFeCo/Pt samples.

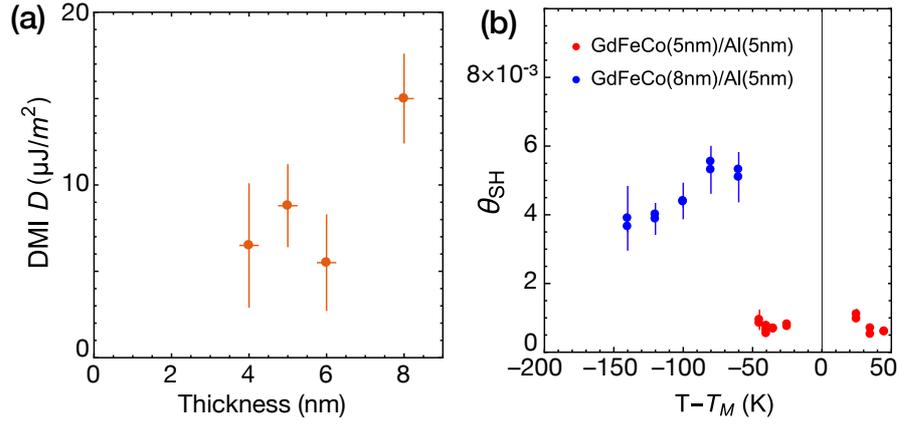

**Figure 5 (a)** The DMI values measured by BLS as a function of GdFeCo thickness (in GdFeCo t nm /Al 5 nm) **(b)** Spin Hall angle as a function of temperature for 5 nm GdFeCo/Al (red) and 8nm GdFeCo/Al samples (blue).

## VI. Composition Profile

The presence of DMI and SOT in thin films without an HM adjacent layer is far from trivial and suggests that its origin issues from the intrinsic properties of the magnetic film. These phenomena require a broken inversion symmetry. To investigate the origin of the inversion asymmetry in the film, electron energy-loss spectroscopy (EELS) studies were performed in a scanning transmission electron microscopy (STEM) in cross-sectional views [24,25,27,43]. Typical high-angle annular dark-field (HAADF) and bright-field (BF) STEM images of the SiOx/GdFeCo/Al film are shown in figures 6(a) and (b), respectively. The GdFeCo layer can be clearly identified by the higher contrast in the HAADF image, while both SiOx and AlOx layers are distinguishable by their lower contrasts. The ferrimagnetic layer presents a homogeneous thickness of ca. $5.2 \pm 0.4$ nm, with defined bottom and top interfaces running along the observed film (See Supplemental Material at [30], fig. S16). The BF image presents a clear amorphous contrast in the GdFeCo alloy across the total thin film thickness, which rules out any non-centrosymmetric crystal structure.

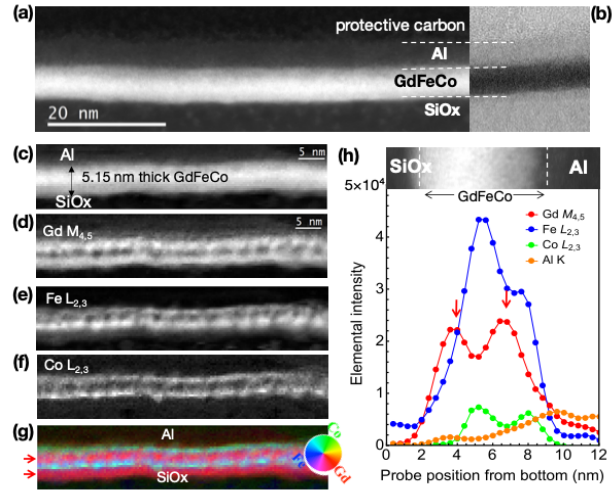

**Figure 6 (a)** HAADF and **(b)** BF STEM images of the SiOx/GdFeCo/Al cross-section. **(c)** HAADF intensity map of the probed area with the corresponding elemental maps extracted at the **(d)** Gd $M_{4,5}$, **(e)** Fe $L_{2,3}$, **(f)** Co $L_{2,3}$ edges. **(g)** Superimposed "false color" map of Gd (red), Fe (blue), and Co (green). **(h)** Laterally integrated line profiles over a length of 12 nm of the Gd $M_{4,5}$ (red), Co $L_{2,3}$ (green), Fe $L_{2,3}$ (blue) and Al K (orange) intensities along the film thickness.

Next, we reveal by EELS a non-uniform elemental distribution in the GdFeCo layer, in agreement with our previous observations in other rare earth/transition metal films [27]. The elemental maps and profiles (Figures 6c to e) clearly evidence an anti-correlated spatial distribution of the Gd (Fig. 6d) and the 3d TM elements (Fe in Fig. 6e and Co in Fig. 6f), throughout the ferrimagnetic layer (See Supplemental Material at [30], fig. S17). This peculiar elemental dissociation is present along the whole amorphous film, while its depth-distribution is nanostructured as follows: a first Gd-rich layer of ca. 1.6 nm at the bottom interface, followed by a first FeCo-rich intermediate layer of ca. 1.2 nm, then a second Gd-rich layer of ca. 1.2 nm, and finally a thin FeCo-rich layer of ca. 1 nm at the top interface. Furthermore, while the Gd-rich layers appear homogeneous along the entire film (marked by red arrows in Figs. 6g and h), the FeCo-rich layers form aggregated nanostructures separated regularly by 2.8 ± 0.4 nm. These findings highlight a nanostructure alternating between Gd- and FeCo-rich sub-layers, in contrast with previous work reported for thicker GdFeCo films [25]. A similar structure was found in the Pt-capped sample (See Supplemental Material at [30]). This elemental inhomogeneity breaks the spatial inversion symmetry and enables the emergence of net SOC effects inside the GdFeCo layer. Similar dependence with composition gradient in heavy metals was reported for FePt film [44] although in TbCo the effects were found to be independent of the composition gradient [45].

# VII. Conclusion

In summary, we have measured a net SOT, a significant DMI, and chiral Néel DWs in single thin ferrimagnetic films without any HM interface. This is a threefold experimental evidence of internal SOC and inversion asymmetry. The fact that DMI and SOT do not decrease with thickness is a further confirmation of dominating internal SOC mechanism. We attribute the SOC to the electronic hybridization of RE 5d and TM 3d electrons. The source of the asymmetry is the elemental inhomogeneity along the depth. The internal DMI and SOTs superimpose with the interfacial effects and, therefore, should not be overlooked in the analysis of stability and dynamics of magnetic textures. Moreover, these internal effects could be engineered by tuning the composition profile along the film thickness. Our findings not only provide insights into the physics of ferrimagnetic alloys but also open additional paths to design and engineer advanced materials for ultrafast spintronics applications.

Note: During the review process, similar internal SOTs were also reported in ferrimagnet [46] and in ferromagnetic CoPt [47], which further supports the explanation based on internal mechanism.

## Data availability

The data are available from the corresponding author upon reasonable request.

## Acknowledgements

We are very thankful to Stanislas Rohart for fruitful discussions, Richard Mattana for the SQUID measurements of $M_S$, and Nathalie Brun for assistance with the analysis of the STEM data. S. K., and E. H. acknowledge public grant overseen by the ANR as part of the "Investissements d'Avenir" program (Labex NanoSaclay, reference: ANR-10-LABX-0035) for the FEMINIST project and travelling grants. S. Krishnia acknowledges a public grant from PIAF ANR-17-CE09-0030. The transport measurements were supported by Université Paris-Sud Grant MRM PMP. We acknowledge funding from the ANR under the "Investissements d'Avenir" program TEMPOS (reference: ANR-10_EQPX-50) for the FIB access.

# Supplementary information

# Spin-orbit coupling in single layer ferrimagnets: direct observation of spin-orbit torques and chiral spin textures

Sachin Krishnia[1,*,†], Eloi Haltz[1,*], Léo Berges[1], Lucia Aballe[2], Michael Foerster[2], Laura Bocher[1], Raphaël Weil[1], André Thiaville[1], João Sampaio[1‡], and Alexandra Mougin[1]

[1]Université Paris-Saclay, CNRS, Laboratoire de Physique des Solides, 91405, Orsay, France.

[2]Alba Synchrotron Light Facility, CELLS, Barcelona, E-08290, Spain.

**Table of contents:**

† sachinbudana@gmail.com

‡ joao.sampaio@u-psud.fr

## S1 Modulation of magnetization due to current-induced effective fields and harmonic Hall voltage expression

The SiOx/GdFeCo/Al sample exhibits tiny internal SOC effects. To obtain accurate and precise SOT fields, we modify the analysis of the harmonic Hall voltage method to accommodate the behavior of $V_{2f}$ with magnetic fields applied in a range between zero to $H >> H_k$. This new approach allows us to deduce the SOT fields with increased accuracy compared with the results obtained with the conventional approach, which uses $H << H_k$ [1].

When an ac current is injected into a heavy metal/magnetic material bi-layer, it induces effective magnetic fields ($\Delta H$) on the magnetization due to various phenomena: Oersted field, or torques due to generated spin currents. The $\Delta H$ induces oscillations of the magnetization around its equilibrium position ($\theta$, $\phi$) in synchronization with ac frequency ($f$) (Figure S1). The magnitude of these oscillations ($\Delta\theta$, $\Delta\phi$) not only depends on the effective field induced by the current, but also on the external field $H$, the anisotropy of the sample $H_k$, etc. By measuring the variation of ($\Delta\theta$, $\Delta\phi$) as a function of applied fields, it is possible to extract the components of the current-induced effective field $\Delta H$.

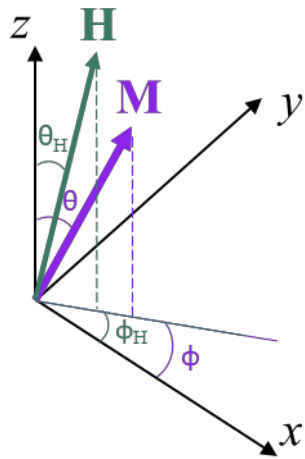

**Figure S1.** A schematic showing the directions of magnetization and external magnetic field in cartesian coordinates system.

Using the Taylor's expansion, $\Delta\theta$, and $\Delta\phi$ can be expressed as:

$$\Delta\theta = \frac{\partial_{H_x} E}{\partial_\theta E} \Delta H_x + \frac{\partial_{H_y} E}{\partial_\theta E} \Delta H_y + \frac{\partial_{H_z} E}{\partial_\theta E} \Delta H_z \quad \& \quad \Delta\phi = \frac{\partial_{H_x} E}{\partial_\phi E} \Delta H_x + \frac{\partial_{H_y} E}{\partial_\phi E} \Delta H_y + \frac{\partial_{H_z} E}{\partial_\phi E} \Delta H_z, \quad (1)$$

with $E$ is the total energy density of the thin film. For a film with uniaxial anisotropy, the energy density can be expressed as:

$$E = -K_{eff}\cos^2\theta - M_S(\sin\theta\cos\phi\, H_x + \sin\theta\sin\phi\, H_y + \cos\theta\, H_z), \quad (2)$$

with $K_{eff}$ the effective out-of-plane anisotropy (given by $K_{eff} = \mu_0 H_k M_S/2$). By combining equations (1) and (2), we obtain:

$$\Delta\theta = \frac{\cos\theta(\Delta H_x\cos\phi + \Delta H_y\sin\phi) - \Delta H_z\sin\theta}{H_k\cos 2\theta + H\cos(\theta - \theta_H)} \quad \text{and} \quad \Delta\phi = \frac{\Delta H_y\cos\phi - \Delta H_x\sin\phi}{H\sin\theta_H}. \quad (3)$$

Now we will show how to extract the magnitudes of these oscillations from the anomalous Hall voltage. For our *x-y* configuration (Figure S2 a and b in the main text), the anomalous Hall voltage $V_{xy}$ is:

$$V_{xy} = I\, R_0 + I\, R_{AHE}\cos(\theta) + I\, R_{PHE}\sin^2(\theta)\sin(2\phi)), \quad (4)$$

with $R_{PHE}$ the planar Hall effect resistance and $R_{AHE}$ the anomalous Hall effect resistance. $R_0$ is the *x-y* resistance due to imperfect cross geometry, which is independent of the magnetization. By taking into account the time modulation of the current at frequency *f*, we replace: $I \to I_0\sin(2\pi f t)$, $\Delta H \to \Delta H \sin(2\pi f t)$, $\theta \to \theta + \Delta\theta\sin(2\pi f t)$ and $\phi \to \phi + \Delta\phi\sin(2\pi f t)$. $V_{xy}$ becomes:

$$V_{XY} = V_0 + V_f\sin(2\pi f t) + V_{2f}\cos(4\pi f t), \quad (5)$$

with:

$$V_f = (R_0 + R_{AHE}\cos\theta + R_{PHE}\sin^2(\theta)\sin(2\phi))I_0 \quad (6)$$

and $V_0 = V_{2f} = \frac{1}{2}(R_{AHE}\sin\theta - R_{PHE}\sin(2\theta)\sin(2\phi))\Delta\theta I_0 - R_{PHE}\sin^2(\theta)\cos(2\phi)\Delta\phi\, I_0.$ (7)

We note that $V_f$ signal carries the information of magnetization rotation ($\theta$ and $\phi$) due to the external field and anisotropy field of sample. The amplitude of magnetization oscillations due to current-induced torques ($\Delta\theta$ and $\Delta\phi$) appears only in $V_{2f}$. We measure $V_f$ and $V_{2f}$ simultaneously using a lock-in amplifier with respect to sweeping external magnetic field.

The current-induced field has two orthogonal contributions, the damping-like ($H_{DL}$) and field-like ($H_{FL}$), which can be obtained by sweeping the applied field along or transverse to the current direction. To avoid a multidomain state, the external magnetic field *H* is applied slightly off-plane with an angle $\delta\theta_H \approx 9°$ in our experiments. When the applied field is parallel to the current direction (along ± *x-axis*) i.e., $\theta_H = \frac{\pi}{2} + \delta\theta_H$ and $\phi = \phi_H = 0$, by implementing the variations of $\Delta\theta$ and $\Delta\phi$ (eq.3) in (eq.7), $V_{2f}$ becomes:

$$V_{2f}^{DL} = -\frac{1}{2}\sin(\theta)\left(\frac{H_{DL}}{H_k\cos(2\theta) - H\sin(\delta\theta_H - \theta)} + \frac{2H_{FL}\xi\sec(\delta\theta_H)\sin(\theta)}{H}\right)R_{AHE}I_0. \quad (8)$$

In the similar way, when the external field is applied transverse to the current direction (along ± y-axis) i.e., $\theta_H = \frac{\pi}{2} + \delta\theta_H$, $\phi = \phi_H = \frac{\pi}{2}$, by implementing the variations of $\Delta\theta$ and $\Delta\phi$ (eq.3) in (eq.7), $V_{2f}$ becomes:

$$V_{2f}^{FL} = \frac{1}{2}\cos(\theta)\sin(\theta)\left(\frac{H_{FL}}{H_k\cos(2\theta) - H\sin(\delta\theta_H - \theta)} + \frac{2\xi H_{DL}\sec(\delta\theta_H)\sin(\theta)}{H}\right)R_{AHE}I_0, \quad (9)$$

where $\xi = \frac{R_{PHE}}{R_{AHE}}$. Its magnitude can be determined with additional measurements of $V_f = \left(\frac{R_0}{R_{AHE}} + \cos\theta + \xi\sin^2(\theta)\sin(2\phi)\right)R_{AHE}I_0$ at $\phi = 45°$ [1].

The values of $H_k$, $\delta\theta_H$ and $\theta(H)$ can be directly extracted from $V_f$. First, $\theta$ can be directly obtained from $V_f$ (the normalised $V_f$ between +1 and -1) as $\theta = \cos^{-1}(V_f)$. Next, the values of $H_k$ and $\delta\theta_H$ can be extracted from $\cos\theta$ as a function of $H$, by fitting with the Stoner-Wohlfarth model [2],

$$\cos\theta = \frac{f}{6} \pm \frac{1}{6}\sqrt{2f^2 - 18e + \frac{54h_z(1+h_{IP}^2)}{f}} - \frac{h_z}{2}, \quad (10)$$

where $h = \frac{H}{H_k}$ is the normalised applied field with in-plane component $h_{IP} = h\sin(\theta_H)$ and out-of-plane component $h_z = h\cos(\theta_H)$, $d = 1 - h^2$, $e = d\cos(\frac{1}{3}\cos^{-1}(54h_{IP}^2\frac{h_z^2}{d^3} - 1))$, and $f = \pm\sqrt{9h_z^2 + 6d + 6e}$. Simulated $\theta$ and $V_f$ with respect to $H/H_k$ are shown in Figure S2.

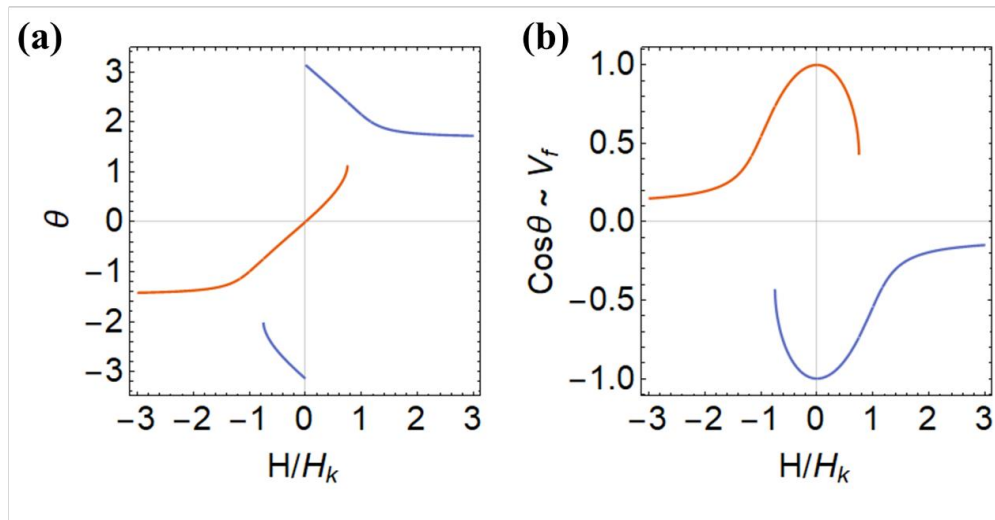

**Figure S2.** Calculated **(a)** magnetization tilting away from z- axis ($\theta$) vs $H/H_k$ and **(b)** First harmonic Hall voltage ($V_f$) vs $H/H_k$ for fields applied $\theta_H \sim 96°$ angle.

To validate the analytical solution, we simulate $V_{2f}^{DL}$ and $V_{2f}^{FL}$ as a function of the external magnetic field. Here, we set $\delta\theta_H = 11°$, and $\xi=0.1$, comparable to our experiments. $V_{2f}^{DL}$ and $V_{2f}^{FL}$ vs $H_x/H_k$ curves for different $H_{FL}$ and $H_{DL}$ as shown in Figure S3 and Figure **S4**. Here we note that the peaks in $V_{2f}^{DL}$ change the polarity when the magnetization is reversed (*i.e.* when *H<0*) whereas $V_{2f}^{FL}$ peaks are independent of magnetization direction. The amplitude of $V_{2f}$ peaks predominantly depends on $H_{DL}$ in longitudinal geometry ($V_{2f}^{DL}$) and on $H_{FL}$ in transverse geometry ($V_{2f}^{FL}$). A similar behavior in the peak polarities and peak heights is observed in our experiments.

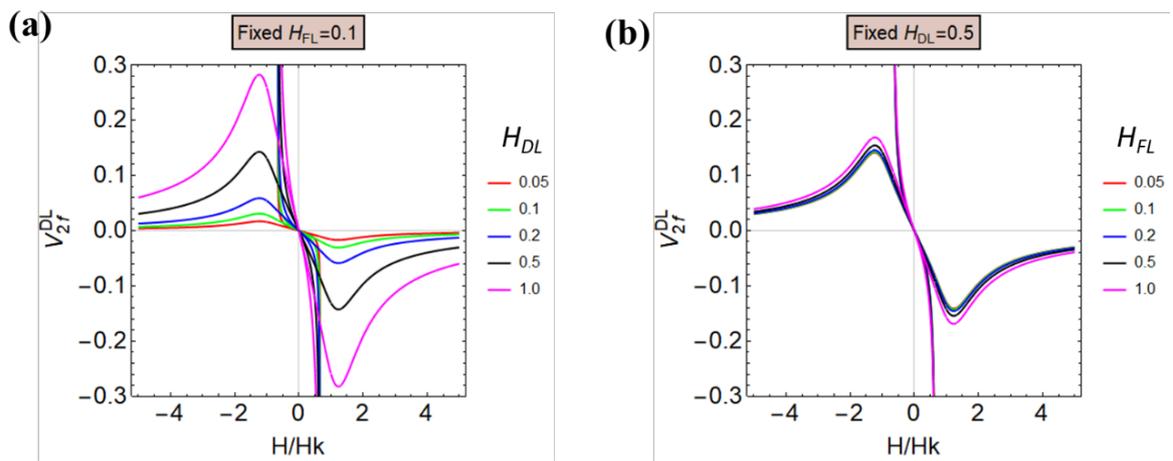

**Figure S3.** Numerically simulated $V_{2f}$ vs in-plane field in longitudinal geometry using equation (8) for $\delta\theta_H = 11°$, and $\xi=0.1$. **(a)** $V_{2f}^{DL}$ vs in-plane field for various $H_{DL}$ at a fixed $H_{FL}=0.1$ and **(b)** for various $H_{FL}$ at a fixed $H_{DL}=0.5$.

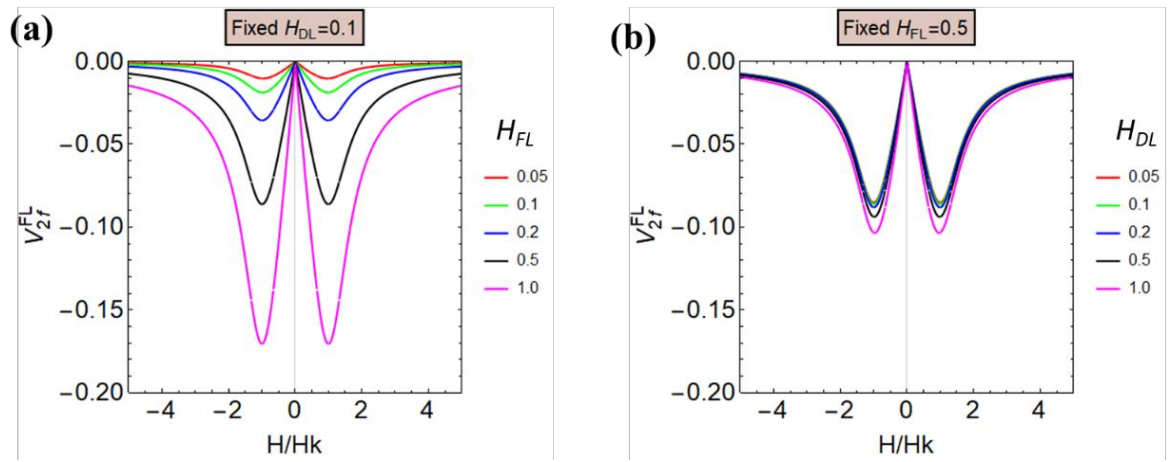

**Figure S4.** Numerically simulated $V_{2f}$ vs in-plane field in transverse geometry using equation (9) for $\delta\theta_H = 11°$, and $\xi=0.1$. **(a)** $V_{2f}^{FL}$ vs in-plane field for various $H_{FL}$ at a fixed $H_{DL}=0.1$ and **(b)** for various $H_{DL}$ at a fixed $H_{FL}=0.5$.

## S2. Spin-orbit torques analysis in SiOx/GdFeCo/Al sample

The 1$^{st}$ ($V_f$) and 2$^{nd}$ ($V_{2f}$) harmonic Hall voltages were measured simultaneously with multiple harmonic lock-in amplifier. So far, various regimes of the $V_{2f}$ vs $H$ plots have been used to extract the effective fields [1,3,4]. Here, we combine all the regimes and perform a full-field fitting method. In other words, we fit the $V_{2f}$ vs $H$ plots from $H=0$ to $H>H_k$. First, we extract $\theta$, $\delta\theta_H$ and $H_k$ from $V_f$ as previously described (eq.10) (Fig.S5(a and c)). Next, by choosing $H_{DL}$ and $H_{FL}$ as free parameters, the experimentally measured $V_{2f}^{DL}$ and $V_{2f}^{FL}$ voltages as a function of the external field are fitted using eq. 8 and eq. 9 (see Figure S5(b) for $H_{DL}$ and (d) for $H_{FL}$). This method allows us to extract $H_{DL}$ and $H_{FL}$ simultaneously from both the geometries. However, the $H_{DL}$ ($H_{FL}$) extracted from longitudinal (transverse) geometry is more accurate as $\xi*H_{FL}$ ($\xi*H_{DL}$) is much smaller than $H_{DL}$ ($H_{FL}$) and has trivial effect on the fit if $\xi \ll 1$. The contribution of $\xi$ is obtained by comparing $V_f$ dependence on in-plane fields applied at azimuthal angle $\phi_H = 0°$ and $\phi_H = 45°$. For the considered sample, $\xi$ is measured to be 0.09. The fittings are performed after subtracting a constant offset from the raw data. We also verify the linear dependence of H$_{DL}$ with the current as shown in Figure S5(e), which overrules significant changes in resistance due to thermal effects. The anomalous Nernst effect, which is proportional to $\nabla T \times m$, is expected to be small as the thermal gradient is perpendicular to the film. It would appear as a hysteretic signal in V$_{2f}$ (see SI of ref. 4), which we either do not observe, or observe to be extremely small (like in Fig S10a). Furthermore, we consider an Oersted field created by the Al cap (assuming 3 nm of the Al cover is oxidized in agreement with the EELS observations). The value of $H_k$ extracted from the $V_f(H)$ fitting are given in the main text Fig 1b.

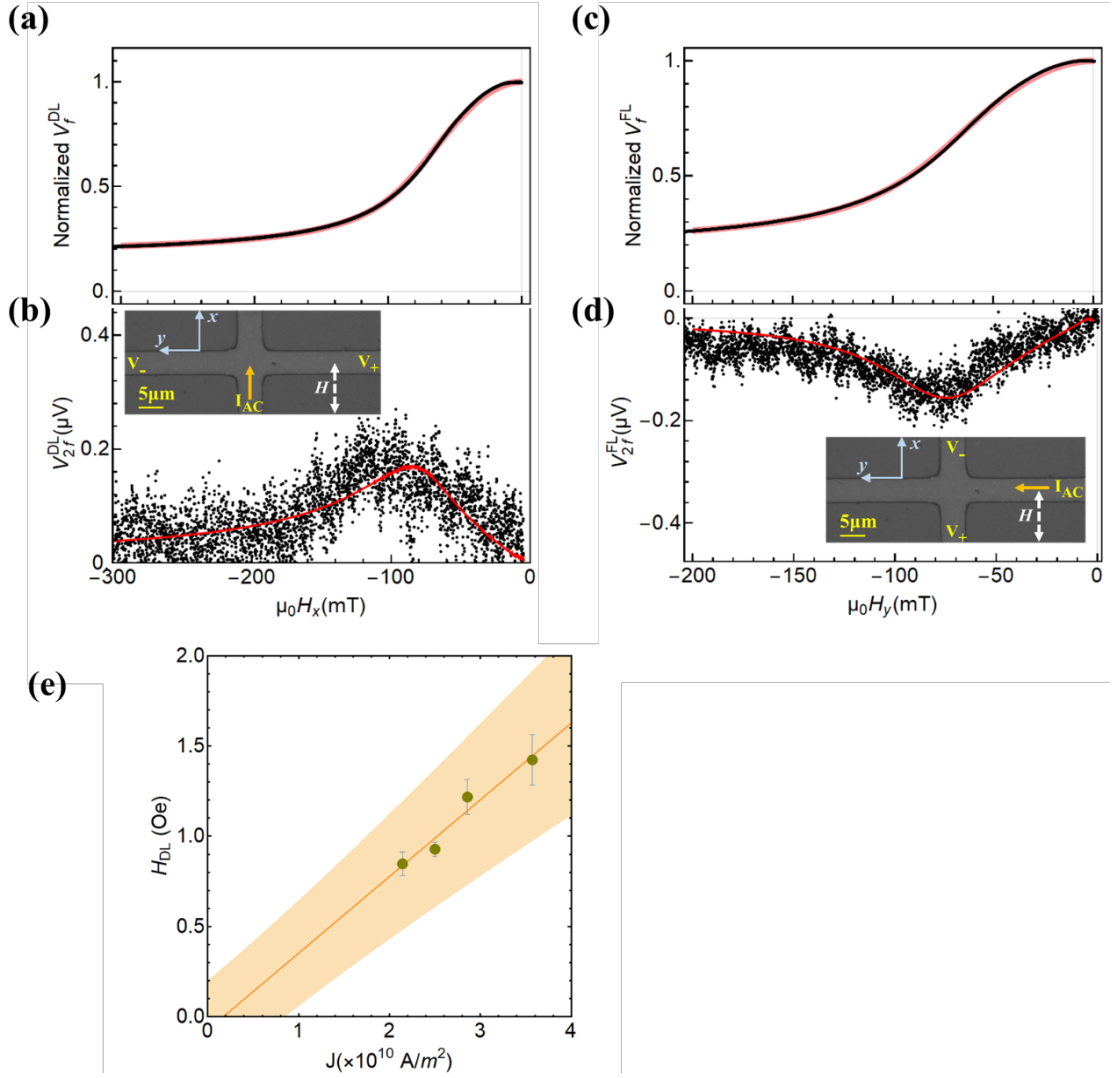

**Figure S5.** **(a)** First and **(b)** second harmonic Hall voltage as a function of external field in longitudinal geometry. **(c)** First and **(d)** second harmonic Hall voltage as a function of external field in transverse geometry. The shown data is measured at $T-T_M=-35K$ and for a current density $1\times10^{10}$ A/m$^2$. The red lines are the fits. Insets are the illustration of measurement technique in longitudinal and transverse geometries. **(e)** Measured damping-like effective field versus applied current density at T-T$_M$= −40 K. The orange envelope is the 95% confidence region of a fit *a x + b* with b ≈ *0*.

Next, the ferrimagnetic thin film goes under spin-flop transition close to T$_M$ where Stoner-Wohlfarth model [2] is not valid to extract the *H$_k$*. As an example, we show below in Figure S6 (a) a measurement of *V$_f$*(H) for the GdFeCo/Pt close to *T*$_M$, where we do observe spin-flop (black points), and the experimental data can not be fitted with Stoner-Wohlfarth macrospin model (pink) [2] whereas it fits very well far from the *T*$_M$ (Figure S6 (b)).

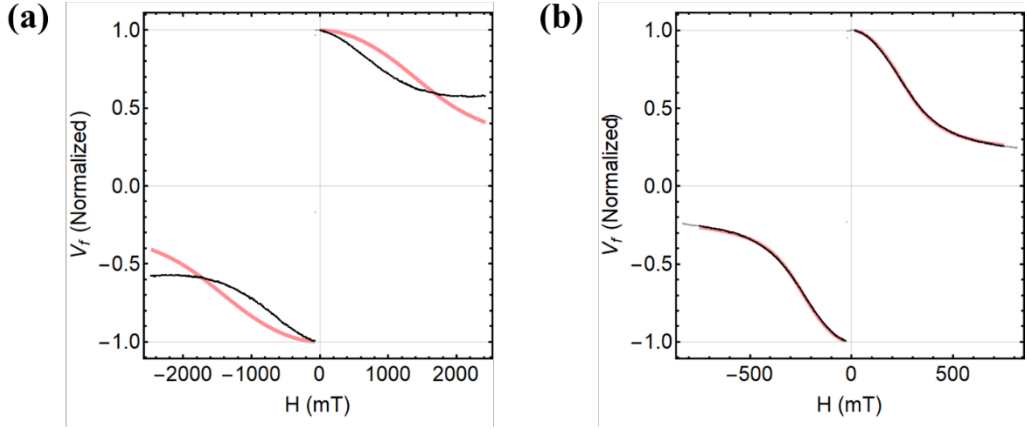

**Figure S6.** Fit (pink lines) of $V_f$ data normalized to range from -1 to +1 (black points) using the Wood 2009 formulas for the Stoner-Wohlfarth macrospin model. This model has only two parameters: the anisotropy field $H_k$ (fitted) and the angle between the field and the film normal (known from other measurements, ~0.15 rad). The conditions on the left (field range, temperature, …) show spin-flop (clear from the inversion of the slope at high field), and lead to a very bad fit. The conditions on the right show a loop without spin-flop and a good fit.

The minimum spin-flop field (close to magnetic compensation temperature, $T_M$) is ~1.4 T for our GdFeCo sample which increases greatly away from $T_M$. Therefore, the torque measurements that are shown in the manuscript were made with much lower field and ~25 K away from $T_M$.

Next, we have compared the results of the fitting by employing the J. Kim's fitting method [1]. This method relies on small angle approximation of net magnetization *i.e.*, the magnetization remains close to the easy axis ($\theta<<1$). With this approximation, the effective fields can be extracted from $V_f$ and $V_{2f}$ data using a relation

$$\Delta H_x \approx \frac{\left(\frac{\partial V_{2f}^{DL}}{\partial H_x}\right)}{\left(\frac{\partial^2 V_f^{DL}}{\partial H_x^2}\right)} \quad \text{and} \quad \Delta H_y \approx \frac{\left(\frac{\partial V_{2f}^{FL}}{\partial H_y}\right)}{\left(\frac{\partial^2 V_f^{FL}}{\partial H_y^2}\right)}, \quad (11)$$

with $H_{DL} = -2\frac{\Delta H_x + 2\xi \Delta H_y}{1-4\xi^2}$ and $H_{FL} = -2\frac{\Delta H_y + 2\xi \Delta H_x}{1-4\xi^2}$. An example of various fittings is shown in Figure S7. The blue fitting is Kim's method, assuming a negligible $\xi$. The red and orange fittings are our analysis methods considering that the longitudinal geometry is sensitive to both $H_{DL}$ & $H_{FL}$ (red) and only to the $H_{DL}$ *i.e.*, $H_{FL}=0$ (orange). The red and orange fittings are quite similar (and the curves are almost perfectly superposed) as expected in the case of no misalignment between current and magnetic field direction and explained in section S2. All the methods give small but significant effective fields within the error bars as shown in Figure S8. However, the values extracted using Kim's method are noisier.

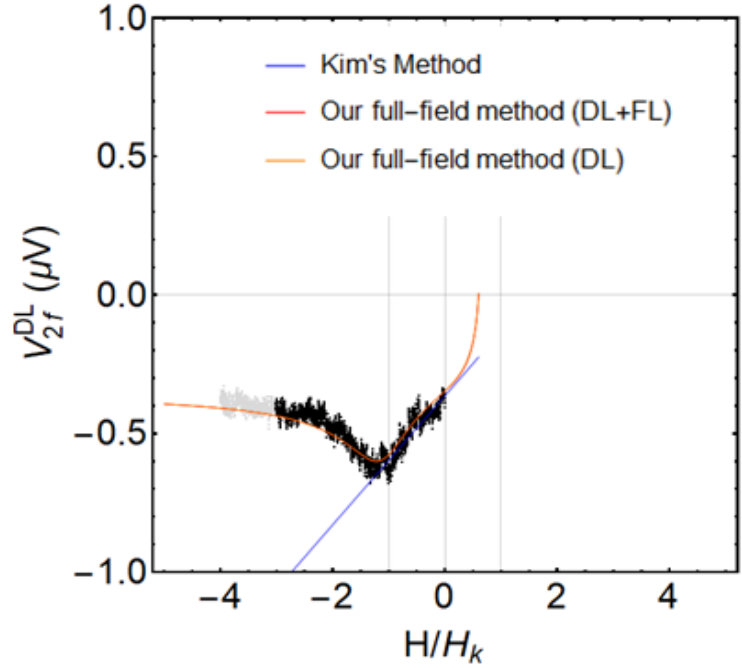

**Figure S7.** Second harmonic Hall voltage as a function of external field in longitudinal geometry. Our full-field fitting method considering the existence of both $H_{DL}$ & $H_{FL}$, and considering that the longitudinal geometry is only sensitive to the $H_{DL}$, i.e., $H_{FL}=0$ are shown by red and orange, respectively (the lines are almost perfectly superposed). Kim's method is shown in blue.

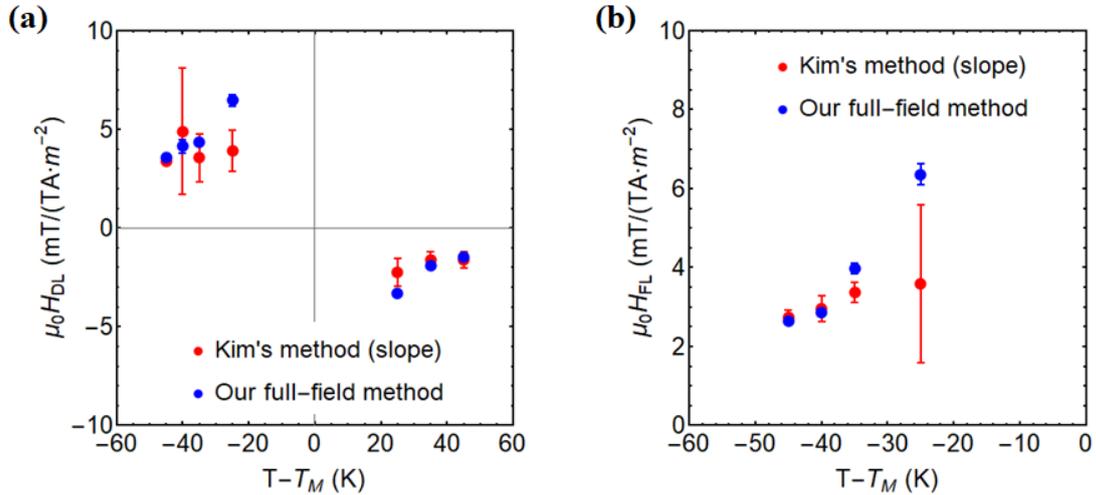

**Figure S8.** Comparison between two analysis methods. **(a)** $H_{DL}$ and **(b)** $H_{FL}$ effective fields per current density as a function of $T$-$T_M$, extracted using our method (blue) and Kim's method (red)

As mentioned in the main text, we could not measure $H_{FL}$ in transverse geometry due to noisy data above room temperature. The raw data for all the temperatures is shown in Figure S9. During the experiments, we interchanged the current and voltage channels instead of rotating the sample, due to the experimental set-up restrictions. The two channels had very different resistance, 3.9 kΩ in longitudinal geometry and 6.5 kΩ in transverse geometry. The larger resistance of the current channel in transverse geometry might be a possible reason for the increasing noise with temperature. However, the exact reason is unknown.

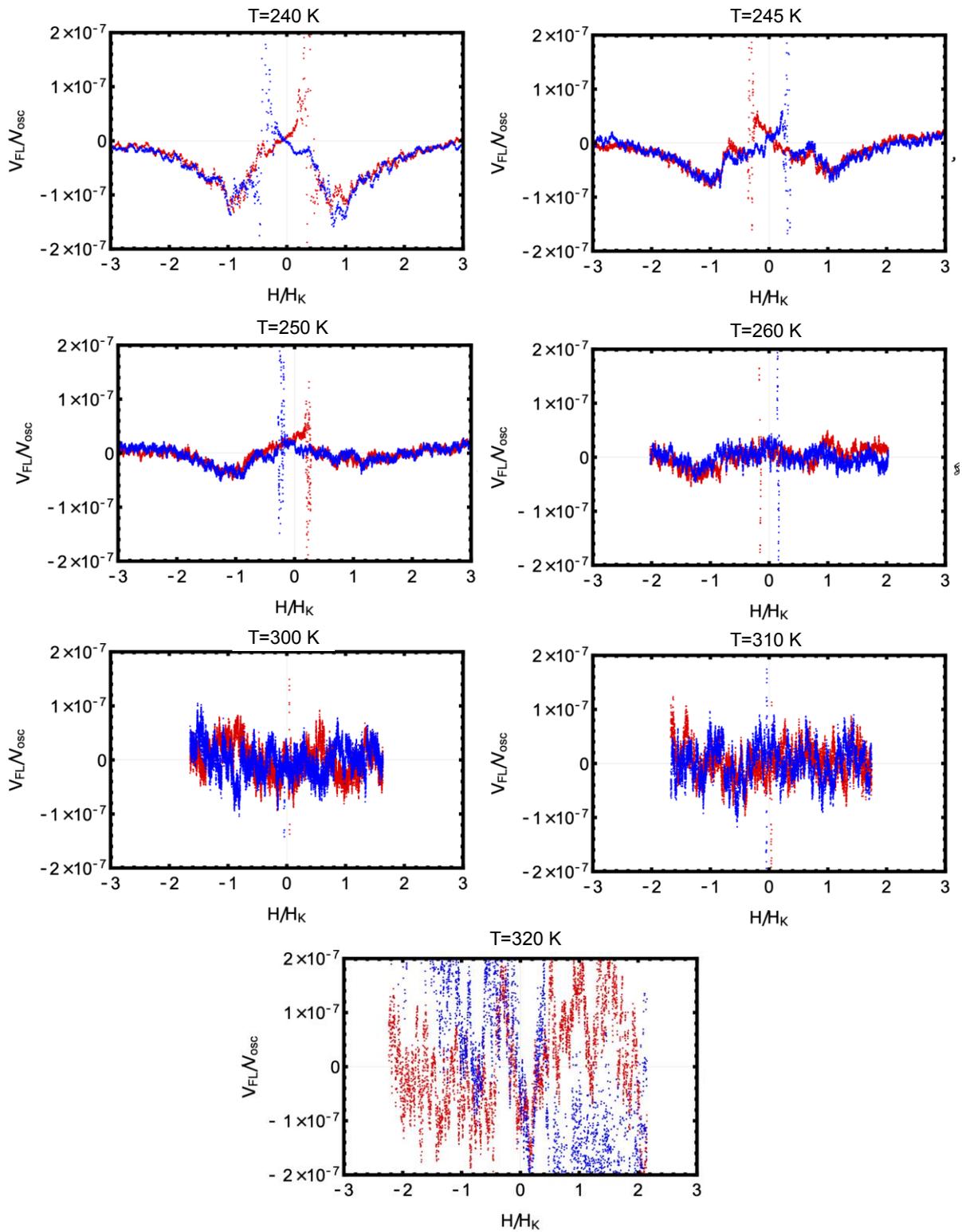

**Figure S9.** Second harmonic Hall voltage signal at various temperatures in transverse geometry. The noise increases with the temperature.

## S3. Spin-Orbit torques analysis in SiOx/GdFeCo/ Pt sample

For the Pt capped sample, as the signals are quite larger, we use Kim's analysis method [1]. Figure S10 (a) shows the behavior of $V_f$ (khaki) and $V_{2f}$ (black) with an external magnetic field in longitudinal geometry. Overlaid red (up magnetization) and blue (down magnetization) lines show the data used for the analysis. Here, we again note that the red and blue data points are almost overlapped i.e., the anomalous Nernst effect is negligible in our Pt capped sample [4]. The $V_f$ ($V_{2f}$) shows quadratic (linear) dependence on in-plane fields. The effective fields are extracted using equation (11). $\xi = 0.08$ is obtained. The effective fields are plotted in Figure S10(b).

The $H_{DLs}$ vs temperature behavior is similar to the Al capped sample i.e., the effective fields diverge at $T_M$, where, $M_s \rightarrow 0$. We estimated the spin Hall angle ($\theta_{SH}$) using the relation $\theta_{SH} = \frac{2e}{\hbar} \frac{\mu_0 H_{DL}}{J} M_s t$ and plotted as a function of temperature in Figure S10 (c). The spin Hall angle is found to be independent of temperature with a median value $\theta_{SH} \approx -0.12$. The sign of spin Hall angle is in agreement with SiOx/Co/Pt structures and opposite to that of the SiOx/GdFeCo/Al sample. The magnitude of the $\theta_{SH}$ is about 100 times higher than in SiOx/GdFeCo/Al sample.

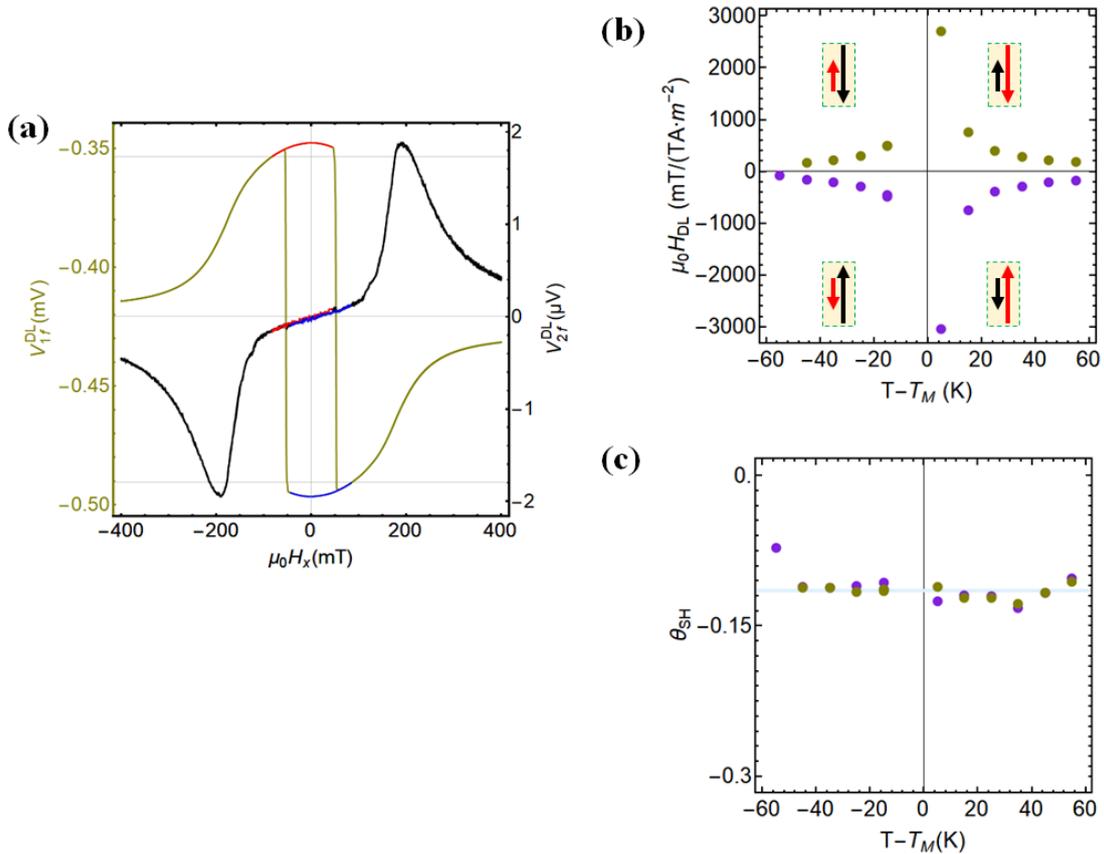

**Figure S10. (a)** First (khaki) and second (black) harmonic Hall voltage as a function of external field in longitudinal geometry for SiOx/GdFeCO/Pt sample below magnetic compensation temperature. **(b)** The $HD_L$ for net magnetization up (purple) and down (khaki) as a function of temperature. The inset shows the direction of Gd (black) and FeCo (red) magnetic moments. **(c)** $\theta_{SH}$ vs temperature. The horizontal blue line corresponds to the median value of $\theta_{SH}$.

Figure S11 (a) shows $V_f$ (khaki) and $V_{2f}$ (black) as a function of external magnetic field in transverse geometry. The $H_{FL}$ is extracted using equation (11) and plotted as a function of temperature in Figure S11 (b). $H_{FL}/J$ also diverges at $T_M$. Sign of the $H_{FL}$ is independent of magnetization direction. If we assume that $H_{FL}$ originates from the Rashba effect, the product $\alpha_R P \approx -2.5 \times 10^{-32} Jm$ is negative and, again much bigger than if compared in SiOx/GdFeCo/Al sample ($\approx 7 \times 10^{-33} Jm$).

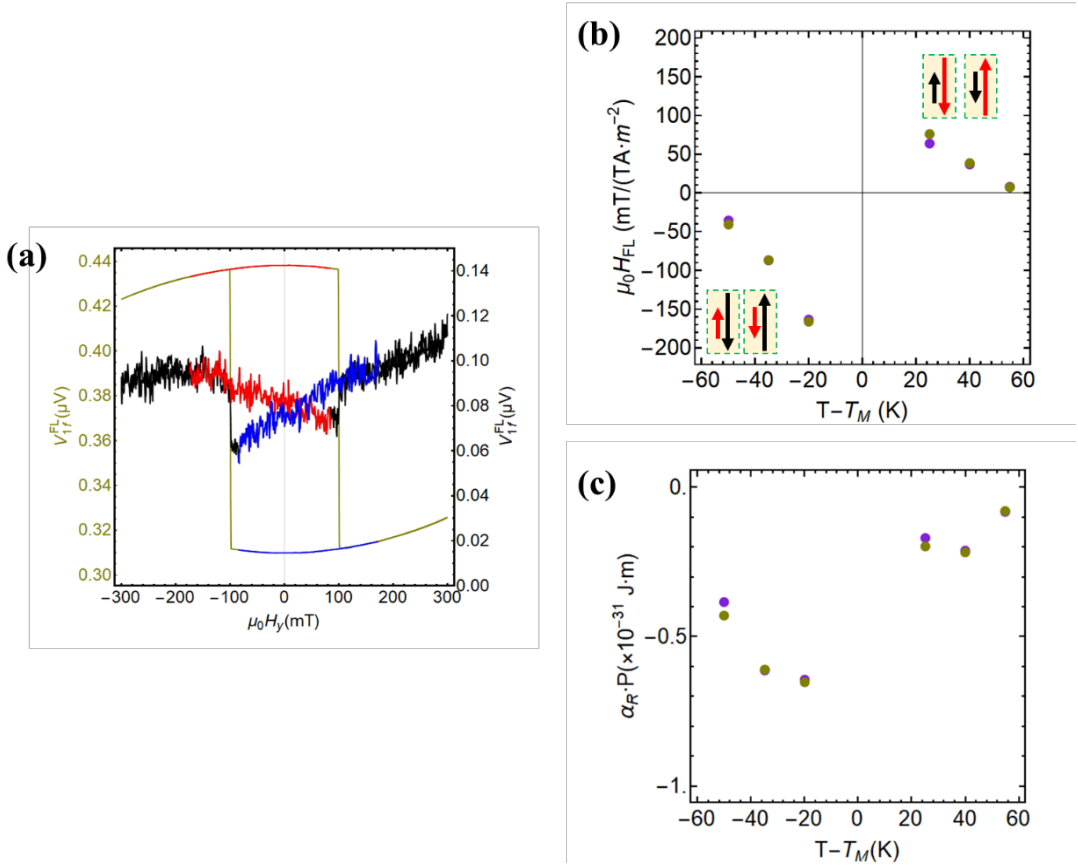

**Figure S11. (a)** First (khaki) and second (black) harmonic Hall voltage as a function of external field in transverse geometry for SiOx/GdFeCo/Pt sample below compensation (T-$T_M$ = − 35K). **(b)** The $H_{FL}$ for net magnetization up (purple) and down (khaki) as a function of temperature. The inset shows the direction of Gd (black) and FeCo (red) magnetic moments. **(c)** $\alpha_R P$ vs temperature.

## S4. XMCD-PEEM imaging on SiOx/GdFeCo/Al sample.

We used X-ray magnetic circular dichroism (XMCD) experiments combined with PEEM at the BL24-CIRCE beamline of the ALBA synchrotron (Barcelona, Spain) to investigate the DWs internal structure in single layer ferrimagnets. The technique uses helicity dependent absorption of circularly polarized X-rays by magnetic moments those are parallel to the beam wave vector ($\vec{k}_b$). Therefore, the image contrast ($I_{XMCD}$) is proportional to the projection of magnetization along the beam direction: $I_{XMCD}(x) = \vec{m} \cdot \vec{k}_b$. The magnetization profile of a DW can be written in polar coordinates as: $m = (sin(\theta) \, cos(\phi), sin(\theta) \, sin(\phi), cos(\theta))$ with $\theta = 2 \, arctan\left[exp\left(\pm\frac{x}{\Delta}\right)\right]$ and $k_{X-Ray} = (sin(\theta_b) \, cos(\phi_b), sin(\theta_b) \, sin(\phi_b), cos(\theta_b))$, which gives:

$$I_{XMCD}(x) = cos(\phi - \phi_b) \, sin\,\theta_b \, sech\left(\frac{x}{\Delta}\right) - cos\,\theta_b \, tanh\left(\frac{x}{\Delta}\right). \tag{12}$$

Figure S12 shows XMCD profiles of an up-down domain wall with right-handed Néel ($\phi = 0$), Bloch ($\phi = \frac{\pi}{2}$) and left-handed Néel ($\phi = \pi$) DWs when $\theta_b = 16°$. The XMCD contrast due to in-plane magnetisation is about ~3.6 times higher than out-of-plane magnetisation for $\phi - \phi_b = 0 \, or \, \pi$. Therefore, the method allows us to probe in-plane & out-of-plane components with high lateral spatial resolution down to 25nm [5].

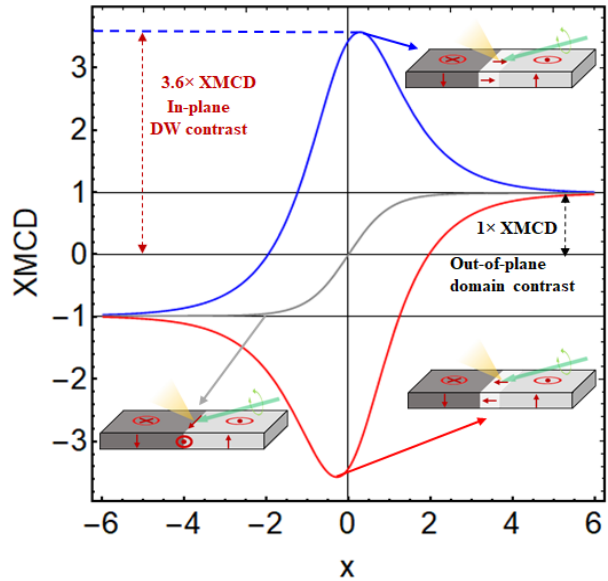

**Figure S12.** Calculated XMCD-PEEM intensity of an Up-Down domain with right-handed Neel ($\phi = 0$), Bloch ($\phi = \pm\frac{\pi}{2}$) and left-handed Neel ($\phi = \pi$) DWs. The DWs magnetization profiles are shown in insets.

Figure S13 shows X-ray absorption spectra (XAS) as a function of photon energy for left-handed and right-handed circularly polarized X-rays. The XMCD signal, *i.e.* ratio of the difference between right and left-handed circularly-polarized beams to their sum, shows a maxima at photon energy 1178.7 eV, which is characteristic of $M_{4,5}$ edge of Gd. In the article, the XMCD-PEEM images are acquired at 1178.7 eV. Note that the chirality of DWs can be determined independently of the XMCD sign.

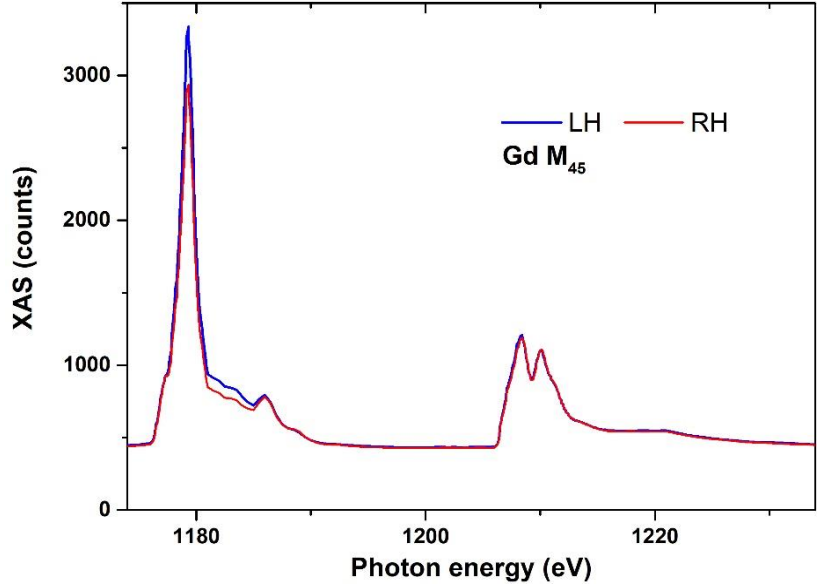

**Figure S13.** X-ray absorption spectroscopy (XAS) for Gd at room temperature. The red and blue curves show the absorption of right-handed (RH) and left-handed (LH) circularly polarized X-rays.

The microscope has a finite resolution *r*. Therefore, we account for the microscope resolution (*r*) by convolving the expected XMCD profile of the DW with a Gaussian function:

$$(I_{XMCD} \star G_r)(x) \triangleq cos(\phi - \phi_b)sin(\theta_b) \int_{-\infty}^{\infty} sech\left(\frac{\eta}{\Delta}\right) G_r(x - \eta)d\eta - cos(\theta_b) \int_{-\infty}^{\infty} tanh\left(\frac{\eta}{\Delta}\right) G_r(x - \eta)d\eta. \tag{13}$$

The Néel nature of the DW can be seen directly in the intense bright and dark contrast of the DWs perpendicular to the X-rays. The DWs internal structure can be further analyzed by fitting the line-scans with equation (13) as shown in Figure 3 of the main text. Few additional DWs profiles are shown in Figure S14. A dip or a peak indicates that the DWs magnetization lies along the X-ray (yellow regions), the Néel DW structure. When the X-ray is parallel to the DW length, peak or dip is not observed, again in the agreement with Néel DW profile (no peak and dip is expected when the DW magnetization is perpendicular to the X-ray direction).

The line-scan is also performed at the edge of a gold pad on the same image into Figure 3 of the main text, as shown in Figure S14(d). The XMCD contrast does not show any peak or dip at the edge of the gold pad, verifying that the observation of peak and dip is not an artifact.

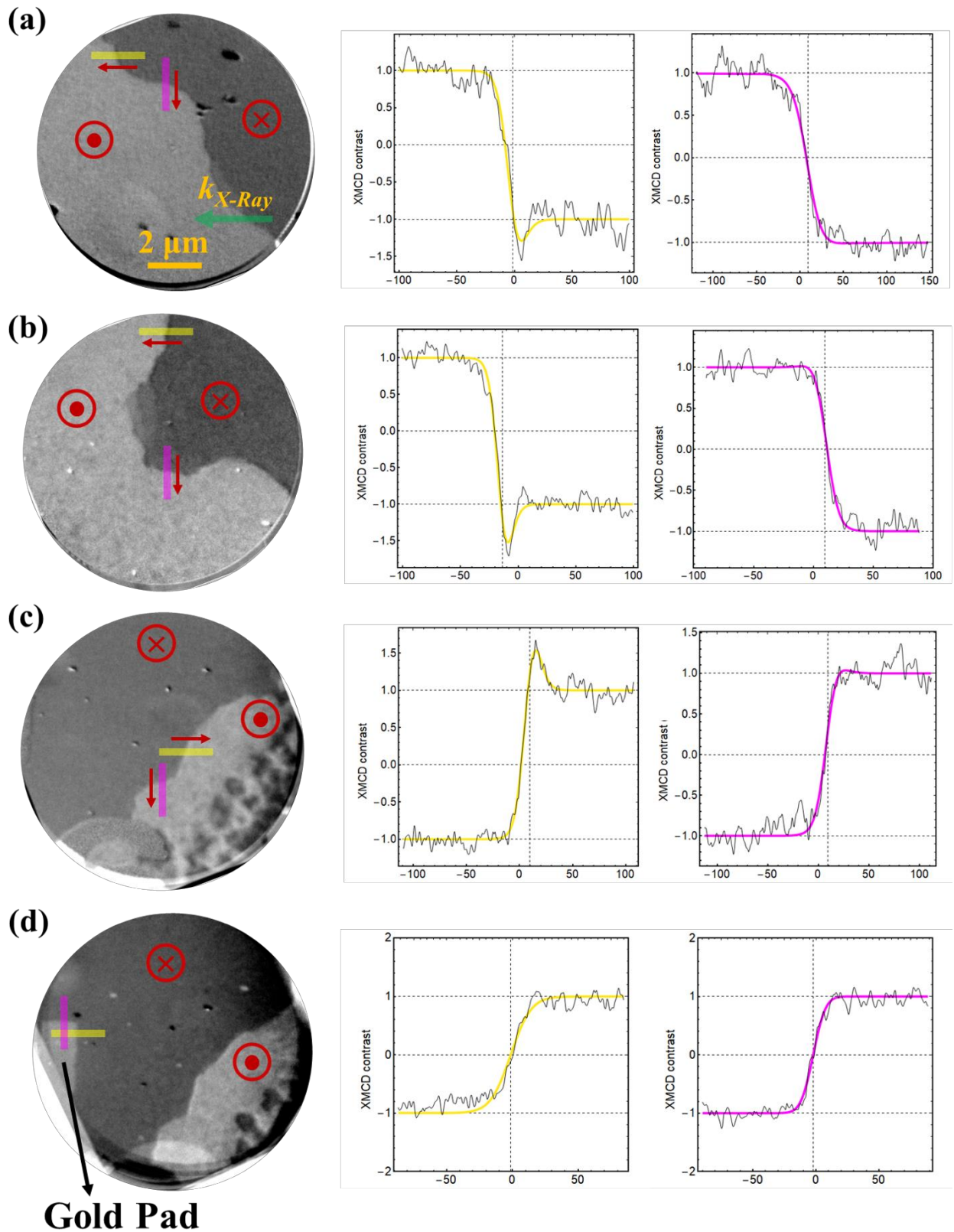

**Figure S14.** Multi-domain XMCD-PEEM images at different parts of the sample at T=230 K. Normalized XMCD intensity profiles (black lines) averaged over the yellow and magenta regions are shown next to each XMCD-PEEM image. The green arrow indicates the incident X-Ray direction. The magnetization direction in domains and domain walls are shown in red.

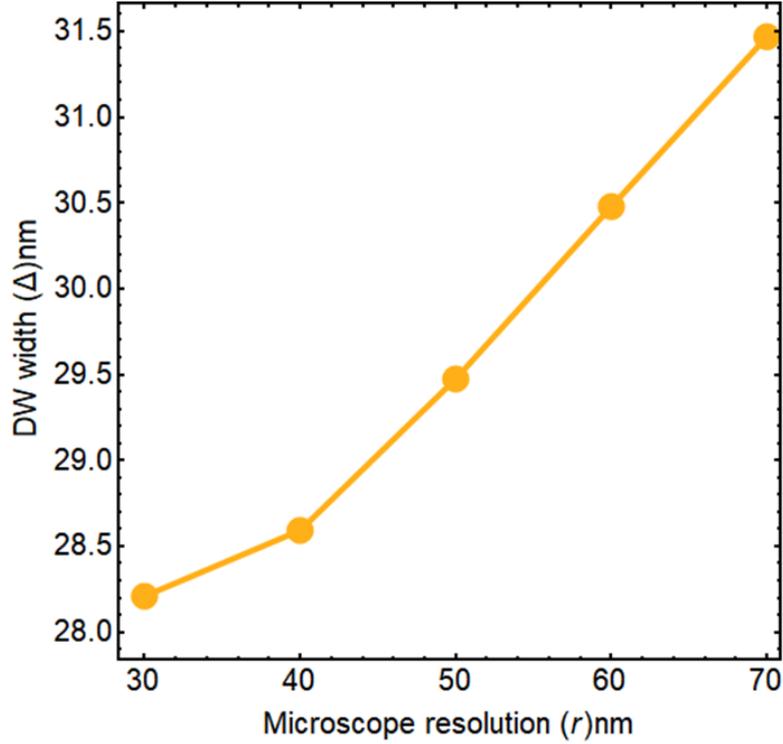

**Figure S15.** The DW width (Δ) as a function of microscope resolution (r).

In the analysis of Néel DWs XMCD intensity profile, the microscope resolution (r) is measured on dust particles on the image. Figure S15 shows the sensitivity of the extracted DW width assuming the r is a fixed parameter with values between 30 to 70 nm. The results show that microscope resolution has a very weak impact on the DW width parameter, Δ≈29±1 nm.

## S5. Spin wave spectroscopy to quantify DMI in SiOx/GdFeCo/Al samples

Brillouin light scattering (BLS) technique in Damon-Eshbach (DB) geometry is used to quantify the DMI in our samples. An in-plane magnetic field ($H_{ex}$) is applied to saturate the magnetization along the y-axis and thermally activated spin waves propagating along x-direction are probed using a green laser of λ = 532 nm. The spectra for different wave-vectors ($k_{sw}$ = 4πSinθ/λ correspond to θ = 10°, 20°, 30°, 45° and 60°) are recorded at T=293K in 180° backscattered geometry and fitted with Lorentzian asymmetric function (an example is shown in Figure 4(a) of main text): $I(f) = \frac{A+c(f-f_0)}{(f-f_0)^2+B^2}$. In addition, when the sample signal was very low, a background (e.g. from the detector's dark current) was included in the fit.

The total momentum is conserved in backscattering events of the light which results in Stokes and anti-Stokes peaks in the spectrum. The position of Stokes and anti-Stokes peaks depends on the energies of the forward and backward propagating SWs. If they have the same energy, the Stokes and anti-Stokes peaks should be symmetric in frequency. However, we observed an asymmetry in the resonant frequencies, which arises due to DMI: $f_S = f_0 \pm f_{DMI}$ and $f_{AS} = f_0 \mp f_{DMI}$. The positions of Stokes and anti-Stokes peaks are extracted by fitting the spectra as shown in Fig.4 of the main text.

The spin-wave dispersion relation is given by [6,7]:

$$f_0 = \frac{|f_S| + |f_{AS}|}{2} = \frac{\gamma\mu_0}{2\pi}\sqrt{(H_{ex} + Jk_{sw}^2 + P(k_{sw}t)M_s)(H_{ex} - H_{K_{eff}} + Jk_{sw}^2 - P(k_{sw}t)M_s)}, \quad (14)$$

Where, $f_0$ is the resonant frequency in the absence of DMI, $M_s = |M^{FeCo} - M^{Gd}|$ is the net magnetization, $J = \frac{2A}{\mu_0 M_s}$, $A$ is the exchange stiffness constant, $H_{keff}$ is the effective anisotropy field, $P(k_{sw}t) = 1 - \frac{1-\exp[-k_{sw}t]}{k_{sw}t}$, $\gamma$ is the effective gyromagnetic ratio of the alloy. $f_0$ as a function of $k_{sw}$ is plotted in **Figure S16**. The extrapolation of these curves to $k_{sw} \to 0$ only depends on $\gamma$ and $H_{keff}$ as:

$$f_{0(k_{sw}\to 0)} = f_{FMR} = \frac{\gamma\mu_0}{2\pi}\sqrt{H_{ex}(H_{ex} - H_{keff})}, \quad (15)$$

The fit (**Figure S16**) gives $\frac{\gamma}{2\pi} = 9.95 \pm 0.15 \frac{GHz}{T}$, yielding a g-factor ≈ 0.71 and $\mu_0 H_k = 0.18$ T at T=293 K. Anisotropy field was extracted from a linear fit to FMR frequency *vs* applied field plot and found to be in agreement with that extracted from AHE measurements on devices. The fit for the full range of $k_{sw}$ gives the $A$= 0.83 pJ/m.

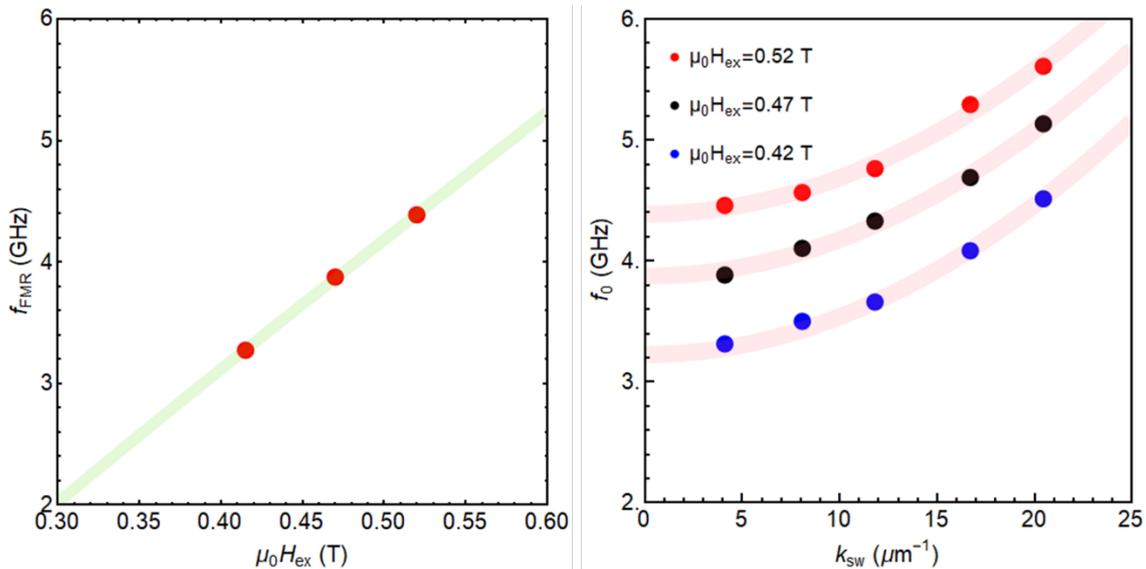

**Figure S16. Left**: Ferromagnetic resonance frequency *vs* applied external magnetic field extracted from the right panel. The line is the fit to the Kittel Eq.15 to extract $H_k$. **Right**: The averaged resonant

frequency ($f_0$) vs $k_{sw}$ for three different in-plane magnetic fields: $\mu_0 H_{ex}$ = 0.42 T (blue), 0.47 T (black), and 0.52 T (red). The lines are the fits to Eq.14 to obtain γ, and $A$.

A relation directly relates the asymmetry in the frequency with DMI constant [8,9]:

$$f_{DMI} = \frac{|f_S|-|f_{AS}|}{2} = \frac{\gamma}{\pi M_s} D k_{sw}. \tag{16}$$

By using the values of γ and $M_s$, the slope of $\Delta f$ vs $k_{sw}$ plot gives the value of DMI constant (Fig.4(b and c) of main text). As the ferrimagnets are very sensitive to the heating effects, the laser power was reduced until no effect of the laser power was detected (0.1 mW).

The difference in the top ($K_{s,top}$) and bottom ($K_{s,bot}$) interfacial anisotropies could also induce Stokes/anti-Stokes frequency shift [10] that could interpret as an apparent DMI ($D_{apparent}$) as below:

$$D_{apparent} = \frac{4}{\pi^2}(K_{s,bot} - K_{s,top})\frac{t^2}{t^2 + (\pi \Lambda)^2}$$

where $\Lambda = \sqrt{2A/(\mu_0 M_s^2)}$. From the DMI value measured in the film of different thicknesses and from the exchange constant determined above by BLS, we estimate $\Lambda \approx$ 22 nm. The fit of this relation to the data yields $K_{s,bot} - K_{s,top} \approx$ 2.8 mJ/m$^2$, an enormous and unrealistic value (the anisotropy in our sample is equivalent to $K_s \approx$ 2 µJ/m$^2$). Therefore, on quantitative arguments, the observed frequency non-reciprocity cannot arise solely from different interface anisotropies.

## S6  STEM and EELS studies on SiOx/GdFeCo/Al sample and SiOx/GdFeCo/Pt samples

Cross-sectional electron-transparent samples were prepared for electron spectro-microscopy (STEM-EELS) studies by focused ion-beam on a SCIOS dual-beam platform (FEI-Thermofischer). The local elemental analysis of the SiOx/GdFeCo/AlOx and SiOx/GdFeCo/Pt nanostructure was performed on a C3/C5-corrected Nion UltraSTEM 200 operated at 100 kV with 30mrad convergence angle, 50mrad EELS collection angle, and around 50-70 pA of probe current. The microscope was equipped with a Medipix3 direct electron detection (Merlin EM Quantum Detectors). EELS spectra were gain-corrected with a gain reference acquired prior to the experiments. Spectrum-images were denoised using a combination of Principal Component Analysis (PCA) (remaining the first 20 components) and denoising of the components maps by BM3D (the state-of-the-art image denoising algorithm [11], followed by an inverse PCA to reconstruct a filtered spectrum-image. The background was fitted with a power-law on an energy window located before the edge and subtracted from the EELS spectra. The absorption edges were integrated over 25 eV for Fe-$L_{2,3}$ and Co-$L_{2,3}$ edges, over 40 eV for Al K and over 60 eV for Gd $M_{4,5}$ edges.

The low-magnification HAADF image of Al cap and Pt cap samples are shown in Figure S17 (a) and S19(a), respectively. The HAADF contrast originates mainly from ~$Z^2$, *i.e.* heavy atoms (e.g. RE and Pt) arise brighter contrasts and lighter atoms (e.g. Al) give darker contrasts. The GdFeCo layers are amorphous and homogenous over several hundreds of nm along the entire film without any apparent defects (Figure S18 for Al cap and Figure S19 for Pt cap sample). At small length scale Gd-rich layers appear homogeneous, while the FeCo-rich layers form aggregated nanostructures separated regularly by 2.8 ± 0.4 nm. Furthermore, they have similar Gd-rich bottom interface (substrate/Gd-rich layer). However, the material of capping layer impacts the top interface. The GdFeCo/Al interface is clean and with slight interdiffusion of Al. In contrast, the Pt diffuses gradually down into amorphous GdFeCo as shown in Figure S19 (g) and (h). Similar intermixing has been observed at top interface of Pt/GdCo/Pt by Quessab et. al [12]. Despite of intermixing and roughness at top GdFeCo/Pt interface, we observe the DMI consistently with Co/Pt interface and opposite in sign to that of GdFeCo/Al sample.

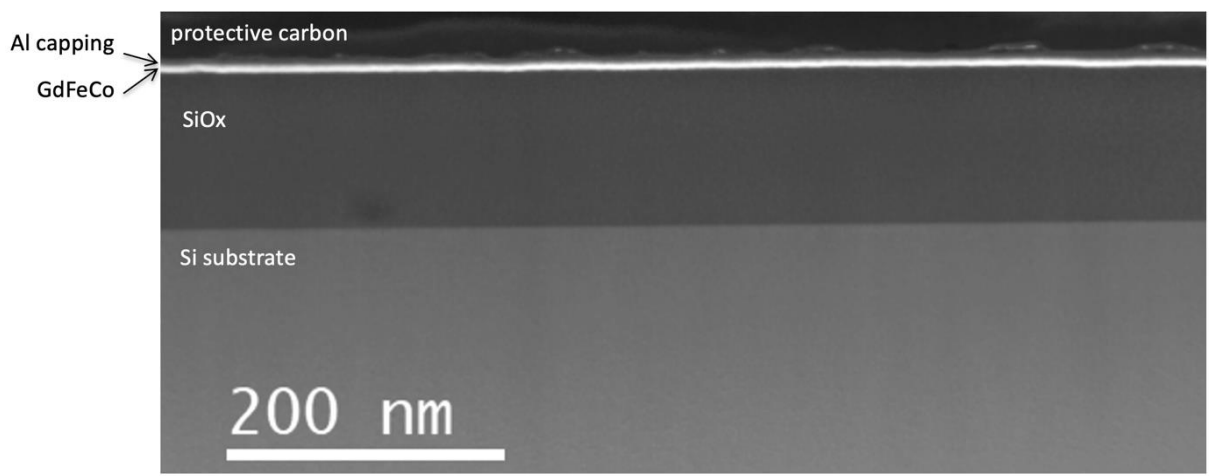

**Figure S17.** Low-magnification HAADF image of the Si/SiOx/GdFeCo/Al sample

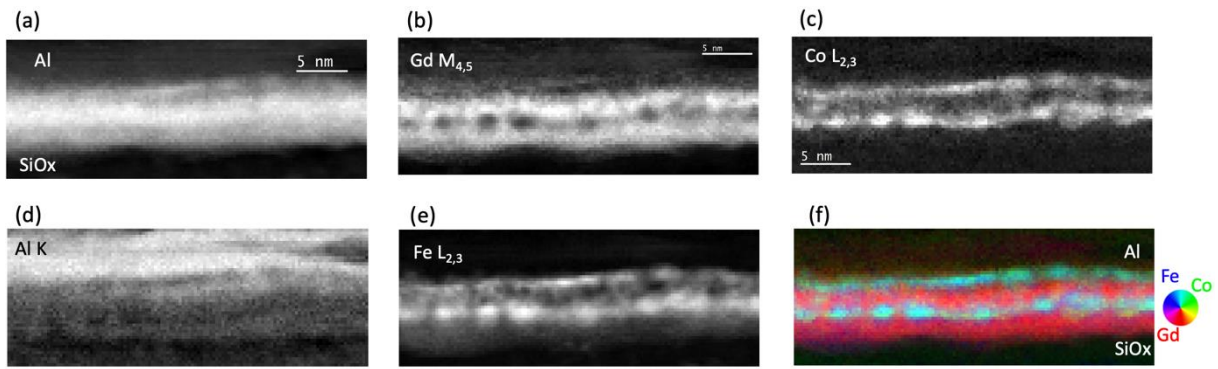

**Figure S18.** Elemental maps of the SiOx/GdFeCo/Al sample: (a) HAADF intensity map acquired simultaneously with the EELS spectrum-image. (b) Gd $M_{4,5}$, (c) Co $L_{2,3}$, (d) Al K, (e) Fe $L_{2,3}$ and (f) superimposed "false color" Gd (red) Fe (blue) and Co (green) maps.

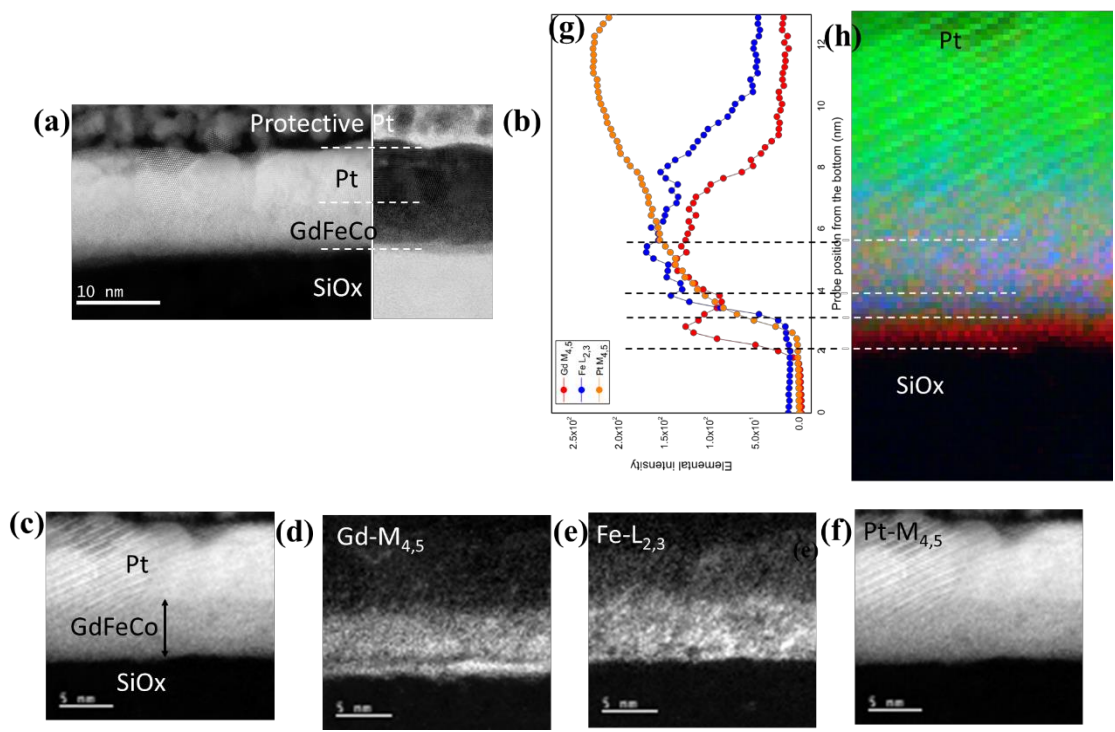

**Figure S19.** Elemental maps of the SiOx/GdFeCo/Pt sample: (a) HAADF and (b) BF STEM images of the SiOx/GdFeCo/Pt cross-section. (c) HAADF intensity map of the probed area with the corresponding elemental maps extracted at the (d) Gd $M_{4,5}$, (e) Fe $L_{2,3}$, (f) Pt $M_{4,5}$ edges. (g) Superimposed "false color" map of Gd (red), Fe (blue), and Pt (green). (h) Laterally integrated line profiles over a length of 12 nm of the Gd $M_{4,5}$ (red), Fe $L_{2,3}$ (Blue) and Pt $M_{4,5}$ (Orange) intensities along the film thickness.